\documentclass[aip,amsmath,amssymb,twocolumn,]{revtex4}                % two column

\usepackage{textcase}
\usepackage{graphicx}% Include figure files
\usepackage{dcolumn}% Align table columns on decimal point
\usepackage{bm}% bold math
\usepackage{amsmath,amsthm,amssymb,mathrsfs}

\begin{document}

\title{Mutual synchronization of spin-torque oscillators consisting of perpendicularly magnetized free layers and in-plane magnetized pinned layers}

\author{Tomohiro Taniguchi}
\author{Sumito Tsunegi}
\author{Hitoshi Kubota}

\affiliation{ 
National Institute of Advanced Industrial Science and Technology (AIST), Spintronics Research Center, Tsukuba 305-8568, Japan}

\date{\today}% 

\begin{abstract}
A mutual synchronization of spin-torque oscillators coupled through current injection is studied theoretically. 
Models of electrical coupling in parallel and series circuits are proposed. 
Solving the Landau-Lifshitz-Gilbert equation, 
excitation of in-phase or antiphase synchronization, depending on the ways the oscillators are connected, is found. 
It is also found from both analytical and numerical calculations that the current-frequency relations for both parallel and series circuits 
are the same as that for a single spin-torque oscillator. 
\end{abstract}

\maketitle

% ===================================================================================================================================================================================== %

Spin-torque oscillators have been a fascinating research target in the field of spintronics 
from the viewpoints of both nonlinear science 
and practical applications such as microwave generator and communication devices 
\cite{kiselev03,rippard04,krivorotov05,houssameddine07,slavin09,bertotti09}. 
Above all, the exciting topic in this research field is the synchronization of spin-torque oscillators 
by the magnetic \cite{kaka05,mancoff05,urazhdin10,locatelli15} and/or electrical \cite{rippard05,zhou09,nakada12,tsunegi16,taniguchi17} couplings. 
The synchronization of spin-torque oscillators results in an enhancement of the emission power and an increase of the quality factor of the practical devices. 
In addition, new applications such as brain-inspired computing based on the synchronized spin-torque oscillators are proposed very recently 
\cite{locatelli14,grollier16,kudo17,torrejon17}. 

% ===================================================================================================================================================================================== %

An attractive structure of spin-torque oscillator for practical applications is that 
consisting of a perpendicularly magnetized free layer 
and an in-plane magnetized pinned layer \cite{rippard10,zeng12,kubota13} 
because this type of spin-torque oscillator results in high emission power, narrow linewidth, and wide frequency tunability simultaneously. 
The oscillation properties of this type of spin-torque oscillator, such as the relation between the injected current and the oscillation frequency, 
as a single oscillator have been investigated both experimentally \cite{kubota13} and theoretically \cite{taniguchi13}. 
A possibility to excite a mutual synchronization in this type of spin-torque oscillators, however, has not been investigated yet. 

% ===================================================================================================================================================================================== %

In this letter, a theoretical study on the mutual synchronization of spin-torque oscillators 
consisting of perpendicularly magnetized free layers and in-plane magnetized pinned layers is presented. 
%Motivated by recent works of the injection locking \cite{khalsa15,tsunegi16}, 
We focus on the coupling of spin-torque oscillators through the current injection, 
and develop models of the coupling in the parallel and series circuits. 
Solving the Landau-Lifshitz-Gilbert (LLG) equation numerically, 
we show that two spin-torque oscillators indicate in-phase or antiphase synchronization 
depending on the way the oscillators are connected. 
An analytical theory clarifying the relation between the current, oscillation frequency, and phase difference is also developed. 
Both the numerical and analytical calculations indicate that 
the dependence of oscillation frequency on the current for both the parallel and series circuits are 
identical to that of a single spin-torque oscillator. 

% ===================================================================================================================================================================================== %

% ===================================================================================================================================================================================== %

\begin{figure}%[p]
\centerline{\includegraphics[width=1.0\columnwidth]{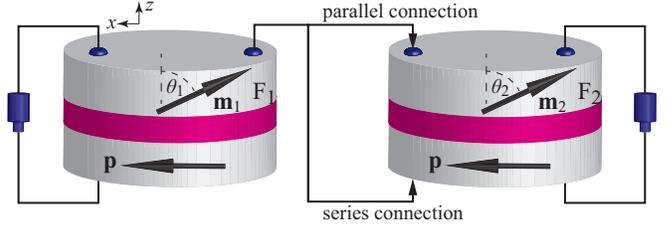}}%\vspace{-3.0ex}
\caption{
        Schematic view of the mutual synchronization of two oscillators coupled through the current injection 
        with parallel or series connection. 
         \vspace{-3ex}}
\label{fig:fig1}
\end{figure}

% ===================================================================================================================================================================================== %

% ===================================================================================================================================================================================== %

The system under consideration is schematically shown in Fig. \ref{fig:fig1}. 
There are two spin-torque oscillators, and each oscillator consists of 
a free layer F${}_{k}$ ($k=1,2$) and a pinned layer. 
For simplicity, we assume that the material parameters of two oscillators are identical. 
The unit vector pointing in the magnetization direction of the F${}_{k}$ layer is 
$\mathbf{m}_{k}=(m_{kx},m_{ky},m_{kz})$, 
whereas the magnetizations in the pinned layers point to the positive $x$-direction, $\mathbf{p}=+\mathbf{e}_{x}$. 
The $z$-axis is normal to the film-plane. 
The external field $H_{\rm appl}$ is applied along the $z$-direction. 
The free layers are perpendicularly magnetized, and therefore, 
the magnetic field acting of the F${}_{k}$ layer is $\mathbf{H}_{k}=[H_{\rm appl}+(H_{\rm K}-4\pi M)m_{kz}]\mathbf{e}_{z}$, 
where $H_{\rm K}$ and $4\pi M$ are the crystalline anisotropy field 
and the shape anisotropy (demagnetization) field along the $z$-direction, respectively. 
%Note that the magnetic field is related to the magnetic energy density via $E=-M \int d \mathbf{m}_{k}\cdot\mathbf{H}_{k}$, %=-MH_{\rm appl}m_{kz}-[M(H_{\rm K}-4\pi M)/2]m_{kz}^{2}$, 
%where $M$ is the saturation magnetization. 
The electric currents are injected to the oscillators and they excite self-oscillations, 
where the positive electric current corresponds to the electrons flowing from the free layer to the pinned layer. 
The magnetization dynamics in the F${}_{k}$ ($k=1,2$) layer is described by the LLG equation, 
\begin{equation}
  \frac{d \mathbf{m}_{k}}{dt}
  =
  -\gamma
  \mathbf{m}_{k}
  \times
  \mathbf{H}_{k}
  -
  \gamma
  H_{{\rm s}k}
  \mathbf{m}_{k}
  \times
  \left(
    \mathbf{p}
    \times
    \mathbf{m}_{k}
  \right)
  +
  \alpha
  \mathbf{m}_{k}
  \times
  \frac{d \mathbf{m}_{k}}{dt},
  \label{eq:LLG}
\end{equation}
where $\gamma$ and $\alpha$ are the gyromagnetic ratio and the Gilbert damping constant, respectively. 
The spin torque strength is given by \cite{slonczewski02} 
\begin{equation}
  H_{{\rm s}k}
  =
  \frac{\hbar \eta I_{k}}{2e (1+ \lambda \mathbf{m}_{k}\cdot\mathbf{p})MV},
  \label{eq:H_s}
\end{equation}
where $M$ and $V$ are the saturation magnetization and volume of the free layer, respectively. 
Two dimensionless parameters, $\eta$ and $\lambda$, determine 
the magnitude and the angular dependence of the spin torque. 
The total current injected into the free layer is denoted as $I_{k}$. 
The explicit form of $I_{k}$ will be given below. 

% ===================================================================================================================================================================================== %

%We consider that two oscillators are connected electrically. 
The spin-torque oscillator generates an oscillating power (current) through the oscillation of the magnetization, 
which can be separated from an external voltage by using bias-Tee \cite{kubota13}. 
The electric current ejected from the spin-torque oscillator, 
which is proportional to $V_{\rm e}/R \simeq \{ V_{\rm e}/[(R_{\rm P}+R_{\rm AP})/2] \}[1 + \Delta R \mathbf{m}_{k}\cdot\mathbf{p}/(R_{\rm P}+R_{\rm AP})]$, 
depends on the magnetization direction through the term $\mathbf{m}_{k}\cdot\mathbf{p}$, 
where $V_{e}$, $R_{\rm P}$, and $R_{\rm AP}=\Delta R + R_{\rm P}$ are the external voltage 
and the resistances at the parallel (P) and antiparallel (AP) alignments of the magnetizations, $\mathbf{m}_{k}$ and $\mathbf{p}$, respectively. 
It has been recently shown both theoretically \cite{khalsa15} and experimentally \cite{tsunegi16} that 
self-synchronization is excited in a vortex oscillator by re-injecting the generated oscillating current into the spin-torque oscillator. 
In this work, on the other hand, the current ejected from the F${}_{k^{\prime}}$ layer is injected into 
the other ferromagnet F${}_{k}$ ($k \neq k^{\prime}$). 
This current excites an additional spin torque on the magnetization $\mathbf{m}_{k}$ in the F${}_{k}$ layer. 
Since the magnitude of this additional spin torque depends on the magnetization direction of the F${}_{k^{\prime}}$ layer, 
the dynamics of $\mathbf{m}_{k^{\prime}}$ influences that of $\mathbf{m}_{k}$. 
Therefore, coupled dynamics of the magnetizations is expected. 

% ===================================================================================================================================================================================== %

% ===================================================================================================================================================================================== %

% ===================================================================================================================================================================================== %

\begin{figure}%[p]
\centerline{\includegraphics[width=1.0\columnwidth]{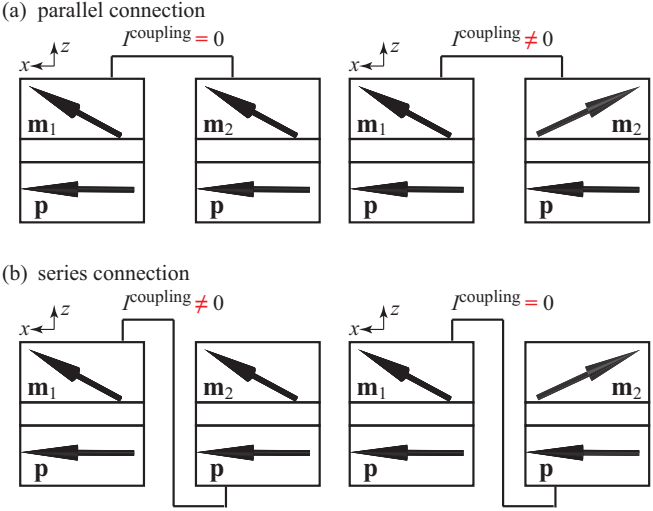}}%\vspace{-3.0ex}
\caption{
        In the parallel circuit, the current flowing through the connection, $I^{\rm coupling}$, becomes zero when $m_{1x}=m_{2x}$, 
        whereas it becomes finite for $m_{1x} \neq m_{2x}$, as schematically shown in (a). 
        On the other hand, in the series circuit, the current in the connection becomes zero when $m_{1x}=-m_{2x}$, 
        as shown in (b). 
         \vspace{-3ex}}
\label{fig:fig2}
\end{figure}

% ===================================================================================================================================================================================== %

% ===================================================================================================================================================================================== %

To establish a model that suits this type of coupling, we consider two types of connections, i.e., 
parallel and series circuits, as shown in Fig. \ref{fig:fig1}. 
The difference between the connections is characterized by the current $I_{k}$ in Eq. (\ref{eq:H_s}). 
Let us denote the current in the absence of the coupling as $I_{0}$. 
In the presence of the coupling, the total current $I_{k}$ can be expressed as 
\begin{equation}
  I_{k}
  =
  I_{0}
  +
  I_{k}^{\rm coupling}
  (\mathbf{m}_{k},\mathbf{m}_{k^{\prime}}),
  \label{eq:current}
\end{equation}
where $I_{k}^{\rm coupling}$ is the current injected from the F${}_{k^{\prime}}$ to the F${}_{k}$ layer. 
We assume that $I_{k}^{\rm coupling}$ is given by 
\begin{equation}
  I_{k}^{\rm coupling}(t)
  =
  \begin{cases}
    I_{0} \chi [m_{kx}(t)-m_{k^{\prime}x}(t)] & ({\rm parallel\ circuit}), \\
    I_{0} \chi [m_{kx}(t)+m_{k^{\prime}x}(t)] & ({\rm series\ circuit}),
  \end{cases}
  \label{eq:current_couple}
\end{equation}
where the dimensionless parameter $\chi$ characterizes the strength of the coupling. 
The parameter $\chi$ reflects the energy loss of the current in the cable connecting the oscillators. 
As shown below, the oscillation frequency of the magnetization is of the order of gigahertz, 
which corresponds to the wave length on the order of centimeter. 
We assume that the spin-torque oscillators are connected by a cable much shorter than the wave length. 
In this case, the coupling occurs instantaneously without any time delay nor phase shift. 
Let us explain the physical meaning of Eq. (\ref{eq:current_couple}). 
In the parallel circuit shown in Fig. \ref{fig:fig2}(a), the current flowing through the connection corresponds to 
the difference between the currents ejected from the two ferromagnets. 
Since the current ejected from the F${}_{k}$ layer includes a term proportional to $\mathbf{m}_{k}\cdot\mathbf{p}=m_{kx}$, as mentioned above, 
the current in the connection is given by $I_{k}^{\rm coupling} \propto m_{kx}-m_{k^{\prime}x}$, as shown in Eq. (\ref{eq:current_couple}). 
This current excites the additional spin torque, and causes the coupled dynamics of the magnetizations. 
When $m_{kx}=m_{k^{\prime}x}$, the currents ejected from the two ferromagnets become the same, 
and, as a result, no current flows in the connection, i.e., $I_{k}^{\rm coupling}$ becomes zero. 
On the other hand, in the series circuit shown in Fig. \ref{fig:fig2}(b), 
the total resistance of the circuit is the sum of the resistances of the ferromagnets, 
which are proportional to $m_{kx}$ and $m_{k^{\prime}x}$. 
Therefore, the current flowing through the circuit includes a term proportional to $m_{kx}+m_{k^{\prime}y}$, 
which corresponds to Eq. (\ref{eq:current_couple}). 
The spin torque excited by this current leads to the coupled dynamics. 
When $m_{kx}=-m_{k^{\prime}x}$, the total resistance, as well as the current flowing through the circuit, becomes independent of the magnetization directions. 
Then, the coupling becomes zero. 

% ===================================================================================================================================================================================== %

% ===================================================================================================================================================================================== %

% ===================================================================================================================================================================================== %

\begin{figure}%[p]
\centerline{\includegraphics[width=1.0\columnwidth]{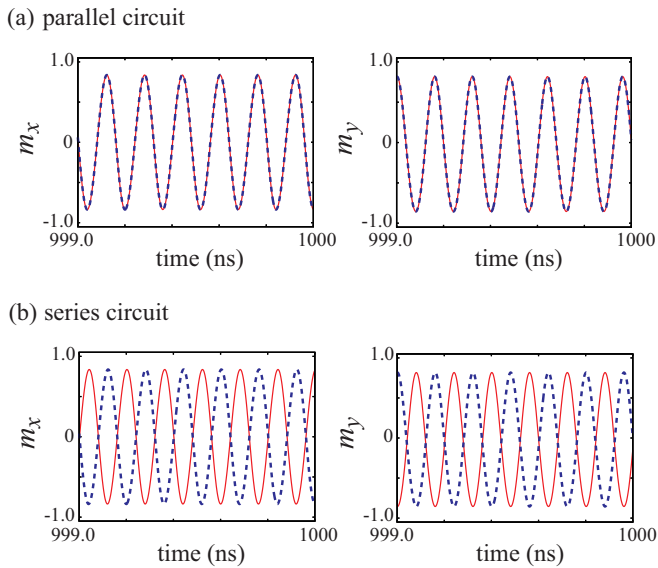}}%\vspace{-3.0ex}
\caption{
        Schematic view of the mutual synchronization of two oscillators coupled through the current injection, 
        where $m_{x}$ and $m_{y}$ of the F${}_{1}$ (red solid) and F${}_{2}$ (blue dotted) are shown. 
        The connections are (a) parallel and (b) series. 
         \vspace{-3ex}}
\label{fig:fig3}
\end{figure}

% ===================================================================================================================================================================================== %

% ===================================================================================================================================================================================== %

% ===================================================================================================================================================================================== %

We study the magnetization dynamics of two spin-torque oscillators by solving Eq. (\ref{eq:LLG}) numerically. 
The values of the parameters are derived from our previous experiment and theory \cite{kubota13,taniguchi13}, 
where $M=1448.3$ emu/c.c., $H_{\rm K}=18.6$ kOe, $H_{\rm appl}=2.0$ kOe, 
$\eta=0.54$, $\lambda=\eta^{2}$, $\gamma=17.64$ Mrad/(Oe s), $\alpha=0.005$, and $V=\pi \times 60 \times 60 \times 2$ nm${}^{3}$. 
The coupling strength $\chi$ is assumed to be $0.1$. 
Figure \ref{fig:fig3}(a) shows the time evolutions of $m_{kx}$ and $m_{ky}$ in a steady oscillation state, 
where the solid (red) and dotted (blue) lines correspond to the F${}_{1}$ and F${}_{2}$ layers, respectively. 
The spin-torque oscillators are coupled through the parallel connection. 
The current is $I_{0}=2.5$ mA. 
Starting from different initial conditions of $\mathbf{m}_{1}$ and $\mathbf{m}_{2}$, 
the dynamics of two magnetizations are gradually synchronized, 
and finally, stabilizes in the in-phase synchronization, i.e., $m_{1x}(t)=m_{2x}(t)$ and $m_{1y}(t)=m_{2y}(t)$. 
On the other hand, the antiphase synchronization, 
$m_{1x}(t)=-m_{2x}(t)$ and $m_{1y}(t)=-m_{2y}(t)$, is stabilized for the series connection, as shown in Fig. \ref{fig:fig3}(b). 
We also notice that the oscillation frequency with a common current is the same for both the parallel and series circuits; 
for example, the frequency is $6.3$ GHz for both circuits in Fig. \ref{fig:fig3}. 
The emission power from an oscillator network is enhanced (reduced) compared to that from a single oscillator 
when the phase difference is in-phase (antiphase). 
Therefore, the in-phase synchronization will be useful to increase the power of the oscillator devices. 
On the other hand, both the in-phase and antiphase synchronizations may be useful for brain-inspired computing such as pattern recognition \cite{maffezzoni15}. 

% ===================================================================================================================================================================================== %

% ===================================================================================================================================================================================== %

% ===================================================================================================================================================================================== %

% ===================================================================================================================================================================================== %

% ===================================================================================================================================================================================== %

We also developed an analytical interpretation of the self-oscillation to verify two important conclusions found in the numerical simulation, 
i.e., the phase difference between the spin-torque oscillators depends on the way the oscillators are connected, 
while the oscillation frequency at a given current $I_{0}$ is independent of the way of connection. 
In terms of zenith and azimuth angles $(\theta_{k},\varphi_{k})$ defined as 
$\mathbf{m}_{k}=(\sin\theta_{k}\cos\varphi_{k},\sin\theta_{k}\sin\varphi_{k},\cos\theta_{k})$, 
the LLG equation (\ref{eq:LLG}) up to the first order of the small parameter $\alpha$ is given by 
\begin{equation}
  \frac{d \theta_{k}}{dt}
  =
  -\gamma
  H_{{\rm s}k}
  \cos\theta_{k}
  \cos\varphi_{k}
  -
  2\pi 
  \alpha 
  f \sin\theta_{k},
  \label{eq:LLG_theta}
\end{equation}
\begin{equation}
  \sin\theta_{k}
  \frac{d\varphi_{k}}{dt}
  =
  2\pi 
  f \sin\theta_{k}
  +
  \gamma 
  H_{{\rm s}k}
  \sin\varphi_{k},
  \label{eq:LLG_varphi}
\end{equation}
where the oscillation frequency $f$ is 
\begin{equation}
  f(\theta)
  =
  \frac{\gamma}{2\pi}
  \left[
    H_{\rm appl}
    +
    \left(
      H_{\rm K}
      -
      4\pi M 
    \right)
    \cos\theta_{k}
  \right].
  \label{eq:frequency}
\end{equation}
The self-oscillation is excited when the spin torque balances the damping torque. 
This condition means that the second and third terms on the right-hand side of Eq. (\ref{eq:LLG}) averaged over an oscillation period cancel each other.
Accordingly, the oscillating magnetizations are mainly described by the first term of Eq. (\ref{eq:LLG}), i.e., the torque due to the magnetic field. 
This torque leads to an oscillation of magnetization on a constant energy curve of $E$, 
where the energy density $E$ is defined as $E=-M \int d\mathbf{m}_{k}\cdot\mathbf{H}_{k}$. 
In the present case, the constant energy curve corresponds to the trajectory with a constant cone angle $\theta_{k}$ in Fig. \ref{fig:fig1} 
because $E=-MH_{\rm appl}\cos\theta_{k}-[M(H_{\rm K}-4\pi M)/2]\cos^{2}\theta_{k}$ depends on $\theta_{k}$ only. 
Since the material parameters of two ferromagnets are assumed to be identical, 
the zenith angle $\theta_{k}$ in the self-oscillation state becomes identical for the two spin-torque oscillators. 
Therefore, in the following, we remove the suffix $k$ from $\theta_{k}$. 
We also call $\theta$ as a cone angle of the oscillation in the following discussion. 

% ===================================================================================================================================================================================== %

Let us first investigate the relation between the phase difference of the spin-torque oscillators and the way the oscillators are connected. 
In the present case, $\varphi_{k}$ can be regarded as a phase of the oscillation. 
According to Eq. (\ref{eq:LLG_varphi}), the phase difference $\Delta\varphi$ defined as $\Delta\varphi=\varphi_{1}-\varphi_{2}$ obeys the following equation, 
\begin{equation}
\begin{split}
  \frac{\sin\theta}{\gamma H_{\rm s0}}
  \frac{d \Delta\varphi}{dt}
  =&
  \frac{[1+\chi(\sin\theta\cos\varphi_{1}\mp\sin\theta\cos\varphi_{2})]\sin\varphi_{1}}{1+\lambda \sin\theta \cos\varphi_{1}}
\\
  &-
  \frac{[1+\chi(\sin\theta\cos\varphi_{2}\mp\sin\theta\cos\varphi_{1})]\sin\varphi_{2}}{1+\lambda \sin\theta \cos\varphi_{2}}, 
  \label{eq:phase_equation}
\end{split}
\end{equation}
where $H_{\rm s0}=\hbar \eta I_{0}/(2eMV)$. 
The double sign $\mp$ means the following: 
the upper ($-$) denotes the parallel circuit and the lower ($+$) denotes the series circuit. 
%It is difficult to solve Eq. (\ref{eq:phase_equation}) for general case. 
%Since the parameters $\chi$ and $\lambda$ are small ($0 \le \chi, \lambda <1$), 
%and 
Since we are interested in the role of the coupling, 
it is natural to consider only the lowest order terms of the coupling strength $\chi$. 
In addition, as the phase $\varphi_{k}$ varies according to $\varphi_{k}=2 \pi f t$, 
we neglect the terms such as $\sin\varphi_{k}$ and $\sin\varphi_{k}\cos\varphi_{k}$, 
which become zero when we focus on an averaged motion of the magnetization during an oscillation. 
Imagine that the phase difference $\Delta\varphi$ slightly shifts from the in-phase ($\Delta\varphi=0$) state as $\Delta\varphi=0+\delta\varphi$. 
Then, the small deviation $\delta\varphi$ from the in-phase state obeys 
\begin{equation}
  \frac{d}{dt}
  \delta
  \varphi
  \sim
  \mp
  \chi 
  \gamma 
  H_{\rm s0}
  \delta
  \varphi. 
  \label{eq:phase_equation_in_phase}
\end{equation}
The solution of Eq. (\ref{eq:phase_equation_in_phase}), $\delta\varphi \propto e^{\mp \chi \gamma H_{\rm s0} t}$, indicates that 
the small deviation in the parallel circuit exponentially decreases with increasing time, 
implying that the in-phase state synchronization is stable in the parallel circuit. 
On the other hand, $\delta\varphi$ in the series circuit increases exponentially, 
indicating that the in-phase synchronization is unstable. 
When we focus on the stability of the phase difference $\Delta\varphi$ 
near the antiphase state ($\Delta\varphi=\pi$), 
we find that a small deviation $\delta\varphi$ from the antiphase state obeys a similar equation to Eq. (\ref{eq:phase_equation_in_phase}). 
However, the double sign $\mp$ in Eq. (\ref{eq:phase_equation_in_phase}) is changed to the opposite sign $\pm$. 
Therefore, in this case, the solution of $\delta\varphi$ exponentially increases (decreases) with increasing time for the parallel (series) circuit, 
indicating that the antiphase synchronization is unstable (stable) in the parallel (series) circuit. 
These conclusions are consistent with the numerical simulations shown in Fig. \ref{fig:fig3}. 

% ===================================================================================================================================================================================== %

% ===================================================================================================================================================================================== %

% ===================================================================================================================================================================================== %

\begin{figure}%[p]
\centerline{\includegraphics[width=1.0\columnwidth]{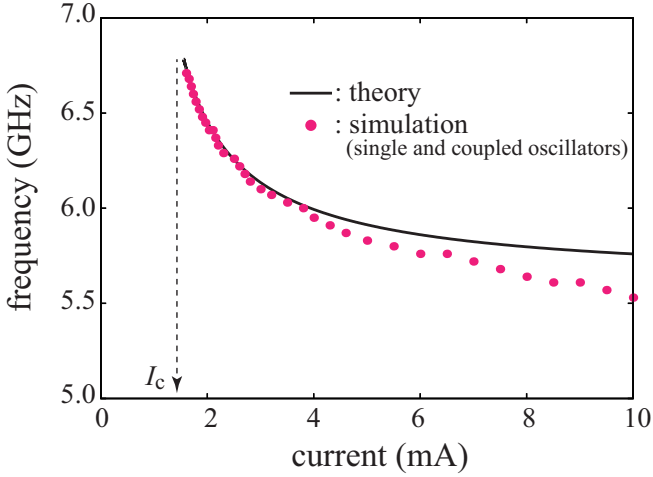}}%\vspace{-3.0ex}
\caption{
         Dependence of the oscillation frequency on the current $I_{0}$ obtained from the numerical simulation (red circles). 
         Although we perform the simulations for three kinds of the spin-torque oscillators, 
         i.e., a single oscillator and synchronized oscillators in both parallel and series circuits, 
         all of the simulations give an identical current-frequency relation. 
         The critical current $I_{\rm c}$ in this study is about 1.6 mA. 
         The black line is obtained from Eqs. (\ref{eq:frequency}) and (\ref{eq:I_theta_sync}). 
         \vspace{-3ex}}
\label{fig:fig4}
\end{figure}

% ===================================================================================================================================================================================== %

% ===================================================================================================================================================================================== %

Next, let us investigate the dependence of the oscillation frequency on the current $I_{0}$. 
As mentioned above, the magnetizations in the self-oscillation state can be approximated as oscillating on a constant energy curve, 
which in the present case corresponds to a trajectory with a constant $\theta$. 
This condition implies that Eq. (\ref{eq:LLG_theta}) averaged over a precession period $\tau=1/f$ is zero, 
i.e., 
\begin{equation}
  \frac{1}{\tau}
  \oint 
  dt 
  \frac{d \theta}{dt}
  =
  0.
  \label{eq:dEdt}
\end{equation}
Using Eq. (\ref{eq:LLG_theta}) with the precession trajectory on a constant energy curve, 
$\mathbf{m}_{1}=(\sin\theta \cos 2\pi f t,\sin\theta \sin 2\pi f t,\cos\theta)$ 
and $\mathbf{m}_{2}=[\sin\theta \cos(2\pi f t-\Delta\varphi), \sin\theta \sin(2\pi f t-\Delta\varphi),\cos\theta]$, 
where $\Delta\varphi$ is the phase difference, 
we find that the current $I_{0}$ satisfying Eq.(\ref{eq:dEdt}) is given by 
\begin{equation}
\begin{split}
  I_{0}(\theta,\Delta\varphi)
  =&
  \frac{2 \alpha e \lambda MV}{\hbar \eta \mathscr{P} \cos\theta}
  \left(
    \frac{1}{\sqrt{1-\lambda^{2}\sin^{2}\theta}}
    -
    1
  \right)^{-1}
\\
  &
  \times
  \left[
    H_{\rm appl}
    +
    \left(
      H_{\rm K}
      -
      4\pi M
    \right)
    \cos\theta
  \right]
  \sin^{2}\theta,
  \label{eq:I_theta}
\end{split}
\end{equation}
where $\mathscr{P}$ stands for 
\begin{equation}
  \mathscr{P}
  =
  \begin{cases}
    1 - \frac{\chi}{\lambda} (1-\cos\Delta\varphi) & ({\rm parallel\ circuit}) \\
    1 - \frac{\chi}{\lambda} (1+\cos\Delta\varphi) & ({\rm series\ circuit}) 
  \end{cases}.
  \label{eq:polarization}
\end{equation}
Since the phase difference $\Delta\varphi$ is zero ($\pi$) for the parallel (series) circuit, 
we notice that Eq. (\ref{eq:I_theta}) becomes 
\begin{equation}
\begin{split}
  I_{0}(\theta)
  =&
  \frac{2 \alpha e \lambda MV}{\hbar \eta\cos\theta}
  \left(
    \frac{1}{\sqrt{1-\lambda^{2}\sin^{2}\theta}}
    -
    1
  \right)^{-1}
\\
  &
  \times
  \left[
    H_{\rm appl}
    +
    \left(
      H_{\rm K}
      -
      4\pi M
    \right)
    \cos\theta
  \right]
  \sin^{2}\theta.
  \label{eq:I_theta_sync}
\end{split}
\end{equation}
Equation (\ref{eq:I_theta_sync}) gives the current $I_{0}$ necessary to excite a self-oscillation on a trajectory with a constant $\theta$, 
where the oscillation frequency is given by Eq. (\ref{eq:frequency}). 
The critical current $I_{\rm c}=\lim_{\theta \to 0}I_{0}(\theta)=[4\alpha eMV/(\hbar \eta \lambda)](H_{\rm appl}+H_{\rm K}-4\pi M)$, 
is the minimum current necessary to excite a self-oscillation. 
We note that Eq. (\ref{eq:I_theta_sync}) explains the reason why the oscillation frequencies found at a fixed $I_{0}$ in the numerical simulation 
are identical for the parallel and series connections; 
it is because Eq. (\ref{eq:I_theta_sync}) is independent of the coupling constant $\chi$.

% ===================================================================================================================================================================================== %

Equation (\ref{eq:I_theta_sync}) predicts another interesting conclusion about the current-frequency relation. 
The oscillation frequencies, as well as the cone angles of the magnetizations, 
of the parallel and series circuits at a given current $I_{0}$ are identical to 
that of a single spin-torque oscillator \cite{taniguchi13} 
because Eq. (\ref{eq:I_theta_sync}) is independent of the coupling constant $\chi$. 
We confirm this prediction for a wide range of the current 
by performing the numerical simulations for the single and synchronized spin-torque oscillators. 
The red circles in Fig. \ref{fig:fig4} shows the dependence of the oscillation frequency on the current obtained from the numerical simulation. 
As expected, we find that the current-frequency relation is identical for three types of the spin-torque oscillators, 
namely, a single oscillator and synchronized oscillators in both parallel and series circuits. 
In addition, the theoretical formulas, Eqs. (\ref{eq:frequency}) and (\ref{eq:I_theta_sync}), shown by the black line in Fig. \ref{fig:fig4} work well 
to reproduce the numerical results, 
which evidently suggests the validity of the analytical formulas. 

% ===================================================================================================================================================================================== %

% ===================================================================================================================================================================================== %

% ===================================================================================================================================================================================== %

In conclusion, 
the mutual synchronization of two spin-torque oscillators 
consisting of perpendicularly magnetized free layer and in-plane magnetized pinned layers was investigated theoretically. 
The physical models for parallel and series connections were proposed. 
The numerical simulation of the Landau-Lifshitz-Gilbert equation revealed that 
in-phase or antiphase synchronization is excited selectively, depending on the ways the spin-torque oscillators are connected. 
It was also shown both numerically and analytically that 
the frequency dependence of two coupled oscillators on the current is identical to that of a single spin-torque oscillator. 

% ===================================================================================================================================================================================== %

The authors are grateful to Takehiko Yorozu and Yoji Kawamura for valuable discussions. 
T.T. is thankful to Satoshi Iba, Aurelie Spiesser, Hiroki Maehara, Ai Emura, and Naka Hasegawa 
for their support and encouragement. 

% ===================================================================================================================================================================================== %

%\bibliography{biblist}% Produces the bibliography via BibTeX.

% ===================================================================================================================================================================================== %

\end{document}